\documentclass[10pt,a4paper,twoside,twocolumn,10pt,A4,twocolumn,final]{IEEEtran}
\usepackage[latin9]{inputenc}
\usepackage{amsmath}
\usepackage{graphicx}

\makeatletter

\pdfpageheight\paperheight
\pdfpagewidth\paperwidth

\@ifundefined{date}{}{\date{}}

\usepackage{ifpdf}

\usepackage{cite}
\usepackage{multirow}
\usepackage{multicol}
\usepackage{blindtext}\usepackage{array}
\usepackage{mdwmath}
\usepackage{mdwtab}
\usepackage{eqparbox}
\usepackage{fixltx2e}
\usepackage{stfloats}

\ifpdf
   
\graphicspath{{./FIG/}}
  \DeclareGraphicsExtensions{.pdf,.jpeg,.png}
 \else
   
  \graphicspath{{./FIG/}}
  \DeclareGraphicsExtensions{.eps}
 \fi
\ifCLASSINFOpdf
  
\else
  
\fi

\makeatother

\begin{document}
\pagenumbering{gobble} 

\title{Collision vs non-Collision Distributed Time Synchronization for Dense
IoT Deployments}

\author{\IEEEauthorblockN{Maria Antonieta Alvarez\IEEEauthorrefmark{1}\IEEEauthorrefmark{2},
Umberto Spagnolini\IEEEauthorrefmark{1}}\\
 \IEEEauthorblockA{\IEEEauthorrefmark{1}Dipartimento di Elettronica, Informazione e
Bioingegneria, Politecnico di Milano, Milan, Italy\\
 \IEEEauthorblockA{\IEEEauthorrefmark{2}Facultad de Ingenier�a de Electricidad y Computaci�n,
Escuela Superior Polit�cnica del Litoral, ESPOL,\\
Campus Gustavo Galindo Km. 30.5 V�a Perimetral, P.O. Box 09-01-5863,
Guayaquil, Ecuador}\\
 Email: \{mariaantonieta.alvarez,umberto.spagnolini\}@polimi.it}}
\maketitle
\begin{abstract}
Massive co-located devices require new paradigms to allow proper network
connectivity. Internet of things (IoT) is the paradigm that offers
a solution for the inter-connectivity of devices, but in dense IoT
networks time synchronization is a critical aspect. Further, the scalability
is another crucial aspect. This paper focuses on synchronization for
uncoordinated dense networks without any external timing reference.
Two synchronization methods are proposed and compared: i) conventional
synchronization that copes with the high density of nodes by frame
collision-avoidance methods (e.g., CSMA/CA) to avoid the superimposition
(or collision) of synchronization signals; and ii) distributed synchronization
that exploits the frames' collision to drive the network to a global
synchronization. 

The distributed synchronization algorithm allows the network to reach
a timing synchronization status based on a common beacon with the
same signature broadcasted by every device. The superimposition of
beacons from all the other devices enables the network synchronization,
rather than preventing it. Numerical analysis evaluates the synchronization
performance based on the convergence time and synchronization dispersion,
both on collision and non-collision scenario, by investigating the
scalability of the network. Results prove that in dense network the
ensemble of signatures provides remarkable improvements of synchronization
performance compared to conventional master-slave reference.
\end{abstract}

\section{Introduction}

The exponential increment of interconnected devices makes crucial
a proper management of cooperative communication and coordinated medium
access control which are enabled by a proper synchronization across
the network. The inter-connectivity of a large number (say $>100$)
of heterogeneous and mutually interfering devices, and their synchronization
are open issues. Internet of things (IoT) \cite{Zanella2014} embeds
the intercommunication of heterogeneous devices in different contexts,
such as smart home, medical metering, public safety, smart grids,
automobile, smart traffic, etc. Factory of Things (FoT) \cite{Savazzi2014}
enables the connectivity for manufacturing with critical latencies.
Even if these paradigms are expected to be part of discussion on 5G
systems \cite{5G}, there are still some characteristics to be considered
for a proper synchronization solution in a dense (says $\gg50-100$
nodes) interconnected network, mainly: i) to allow scalability of
the network as it could be difficult to provide a common synchronization
reference to all node of the network as this would scale poorly with
number of nodes; ii) to allow fast synchronization of the whole network;
and iii) to mitigate power consumption that might follow from excessive
synchronization signaling.

\begin{figure*}[!tp]
\begin{centering}
\includegraphics[width=17cm,height=7cm]{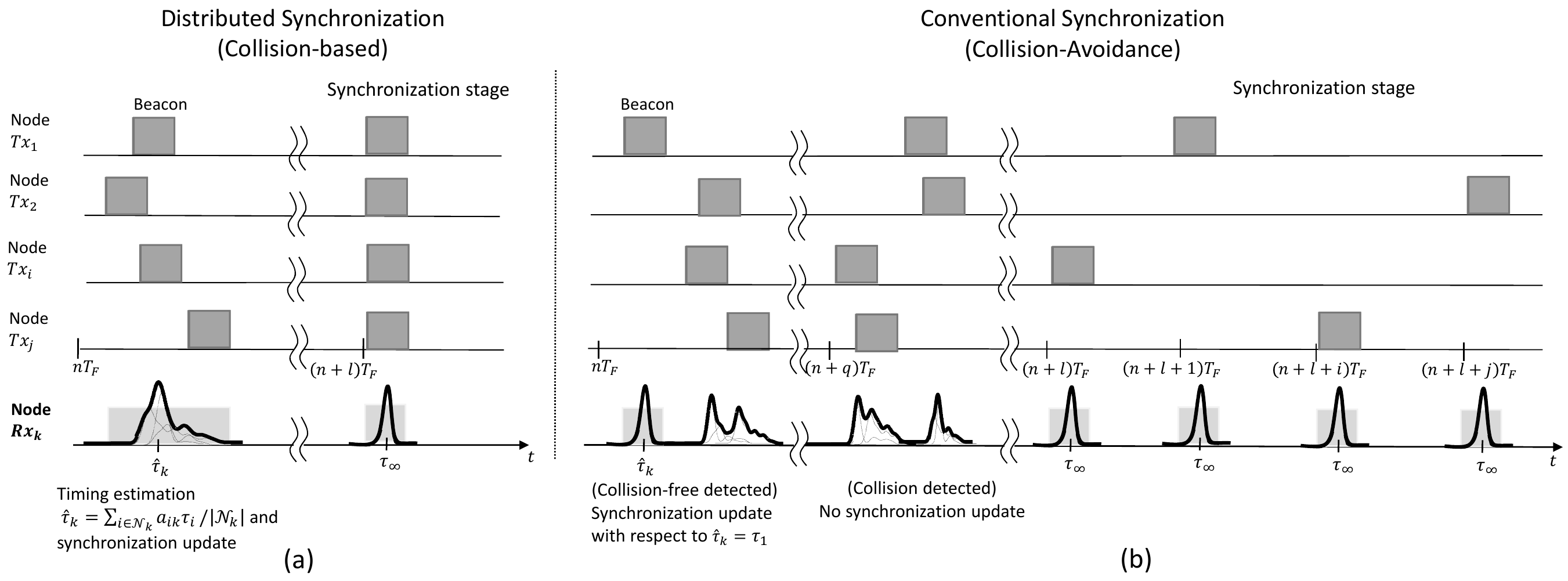}
\par\end{centering}
\caption{Sketch of TO synchronization evolution. (a) Distributed synchronization
based on colliding signals. At received node, say $k$-th, the TO
is updated based on the relative TO error locally estimated with respect
to the ensemble of received signals $\hat{\tau}_{k}=\sum_{i\in\mathcal{N}_{k}}a_{ik}\tau_{i}$
used as reference. At synchronization stage, all frames are TO-aligned
to the start of the frame. (b) Synchronization based on collision-avoidance.
The $k$-th received node searches for signal free-of-collision (left)
to update its TO based on the relative TO error with respect to the
collision-free slot (e.g., $\hat{\tau}_{k}=\tau_{1}$ in (b)) used
as reference. If there is no signal free-of-collision (middle), there
is no correction of the TO. At synchronization, the frame of each
node starts at the same timing with a node-dependent shift.\label{fig:Synchronization-evolution}}
\end{figure*}

In an environment where the closeness and density of nodes makes them
prone to collision of transmitted beacon signals, there has been an
intense research to reduce (or avoid) the collision of beacon signals.
For example, hierarchical architecture or multi-hope scenario by scheduling
the node transmission implying the need of coordination among nodes
\cite{Abedini2016,Xu2016}. IoT technologies have been built based
on several standards that support the connectivity such as IEEE 802.15.4e
\cite{Std802.15.4} with Time-Slotted Channel Hopping (TSCH) that
exploits channel hopping to avoid interference operating at the same
frequency band and low-power consumption. In wirelessHART \cite{Savazzi2014}
synchronization is based on scheduling where the time is divided into
time slots and transmissions within a time slot follow some specific
timing requirements. In IEEE 802.15.4e the network synchronization
is defined on MAC-layer schedule, and it can be centralized (``coordinated''
node is responsible to built and maintain the network schedule) or
distributed (each node decides on which links to schedule with each
neighbor) \cite{Palattella2013}. However, the maintenance of the
schedule is a drawback as the node has to wake up due to lack of scheduling
and this limits the scalability of the network and their synchronization
\cite{Ben2016,Kim2016}.

Thus, instead to avoid the signals' collision, the distributed synchronization
method proposed here exploits the superimposition of transmitted signals
to drive the network to a global synchronization. The distributed
synchronization algorithm is based on consensus paradigm that enables
the network to reach asymptotically a global convergence based on
the exchange of a common beacon (i.e., the same synchronization beacon
is used by all nodes in the network) with features that enable a fast
and accurate timing synchronization. Namely, the colliding signals
from a set of transmitting nodes provide the ensemble reference to
enable the global synchronization of the other nodes that are on receiving
state.

The beacon structure is based on chirp-like signature to give the
start of the frame at the physical layer interface, so that locally
all the devices can estimate and correct their timing offset (TO)
of the frame structure based on distributed phase locked loop (D-PLL)
algorithm \cite{Spagnolini2008}. The devices cooperate with each
other to asymptotically reach a global network synchronization in
only few beacon exchanges without any external coordination, or any
master node acting as TO reference. Once the network reaches a consensus,
the frames are aligned giving the start of time-slot and thus enabling
any MAC-layer protocol such as slotted ALOHA.

This paper compares the synchronization for dense wireless network
(Fig. \ref{fig:Synchronization-evolution}): frame collision-avoidance
(based on conventional synchronization) and frame-collision (based
on distributed synchronization). In order to establish a fair setting
where both synchronization approaches can be compared, we choose a
scenario where each node autonomously and independently decide if
to transmit or receive under half-duplex constraint. The optimum duplexing
strategy is evaluated here for uncoordinated arbitrarily dense network.
Contribution is the analytical derivation of the optimum duplexing
strategy that maximizes the synchronization convergence of the network.
The duplex strategy is addresses in \cite{Alvarez2015} for distributed
synchronization algorithm. Here, it is extended to the conventional
non-collision based synchronization by proving that in this case the
convergence and residual TO error is very sensitive to the choice
of the optimum duplexing and the network density. Numerical analysis
confirms that in dense networks, the collision of beacon-signals provides
an excellent reference for distributed synchronization that implicitly
averages over the active nodes the individual impairments (e.g., clock
drift), and the performance is almost independent on the number of
nodes. This offers a unique feature to scalability for inter-connected
heterogeneous devices as it is in the case of IoT networks.

The paper is organized as follows. Section \ref{sec:System-Model}
defines the system model. The synchronization algorithm based on collision-avoidance
and the distributed synchronization algorithm based on consensus paradigm
are introduced \ref{sec:Colliding-vs-not-Colliding}. Finally, Section
\ref{sec:Numerical-Results} demonstrates numerical results followed
by concluding remarks in Section \ref{sec:Conclusion}.

\section{System Model\label{sec:System-Model}}

Let us consider a dense wireless network of $K$ uncoordinated nodes
(devices) fully connected (i.e., all nodes are allocated in a small
geographic area such that each signal transmitted can be received
by almost all the other nodes, so propagation delay is negligible
and connectivity graph is simple, just for simplicity) without any
external synchronization reference. Time is discretized into frames
and nodes are aware of the nominal frame period $T_{F}$ (this assumption
is not strictly necessary \cite{Spagnolini07}). In network system,
the node timings are nominally the same and oscillators of each node
are affected by independent frequency fluctuations that make the frame-time
$\tau_{k}[n]$, evaluated at $n$-th frame for $t\in[nT_{F}+\tau_{k}[n],(n+1)T_{F}+\tau_{k}[n])$,
and carrier angular frequency $\Omega_{k}(t)=\Omega_{o}+\omega_{k}(t)$
change over time ($\Omega_{o}$ is the nominal RF frequency). Here
it is not considered the analysis for carrier frequency synchronization
to reach $\Omega_{k}(t)=\Omega_{i}(t)$ that would follow a similar
theoretical infrastructure. Similarly, the distributed synchronization
method can be designed for join time and carrier frequency synchronization
as in \cite{Alvarez2014}. The network is synchronized by periodically
exchanging a synchronization signature embedded into a beacon structure.
Each node is equipped with oscillators that run autonomously by their
local drifting, thus timing offset (TO) is locally adapted based on
programmable frequency dividers of local oscillators.

The synchronization signature is denoted as $x(t)$, and the modulated
payload of the $i$-th node within the $n$-th frame is denoted as
$x_{Di}\left(t\right)$. All frames can be considered as mutually
misaligned one another and delayed by the (absolute) time $\tau_{i}[n]$.
Within the $n$-th frame period, each node randomly can be on transmitting
and receiving mode. The transmitting nodes broadcast their synchronization
information, whilst at receiving nodes, they receive the superimposition
of signals to extract the synchronization information and locally
correct their TO. 

The signal received by the $k$-th node over the $n$-th frame interval
$T_{F}$ is the superposition of synchronization signatures $x(t)$
by all the neighboring nodes $\mathcal{N}_{k}$:
\begin{equation}
y_{k}(t|n)=\sum_{i\in\mathcal{N}_{k}}h_{ki}(t)\ast x(t-\tau_{i}[n])\exp(j\Omega_{i}(t)t)+w_{k}(t|n),\label{eq:y(t)}
\end{equation}
here referred to the absolute time-reference, for simplicity. $h_{ki}(t)$
is the channel response for the link $i\rightarrow k$ that accounts
for multi-paths and small propagation delays. Term
\begin{equation}
w_{k}(t|n)=\sum_{i\in\mathcal{N}_{k}}h_{ki}(t)\ast x_{Di}(t-\tau_{i}[n]|n)\exp(j\Omega_{i}(t)t)\label{eq:w(t)}
\end{equation}
is the superposition of all payloads as these are not of interest
for synchronization process, noise can be included as well. This term
is modeled here as white Gaussian as being realistic enough based
on the assumptions so far. The TO error for each of the nodes $\left\{ \tau_{i}\right\} _{i\in\mathcal{N}_{k}}$
belonging to the neighborhood $\mathcal{N}_{k}$ is evaluated by the
$k$-th receiver with respect to the local reference as
\begin{equation}
\Delta\tau_{ki}[n]=\tau_{i}[n]-\tau_{k}[n].
\end{equation}
The TO synchronization is achieved when $|\tau_{i}[n]-\tau_{k}[n]|\leq\sigma_{TO}$,
for any pair $\left(i,k\right)$ with an upper TO limit $\sigma_{TO}$
that depends on the value tolerated by the data-communication protocols.
The relative error value with respect to the reference of the TO of
the $k$-th node is
\begin{equation}
\begin{array}{c}
\Delta\tau_{k}[n]=\hat{\tau}_{k}[n]-\tau_{k}[n],\end{array}\label{eq:relative_sync_par}
\end{equation}
where $\hat{\tau}_{k}\left[n\right]$ is the TO reference. Distributed
synchronization is a consensus-based process where every node is using
as TO reference the average timing $\hat{\tau}_{k}[n]=\sum_{i\in\mathcal{N}_{k}}a_{ik}\left[n\right]\tau_{i}[n]$
($a_{ik}\left[n\right]$ is a weighted coefficient) of the neighboring
nodes that is estimated from superimposed synchronization signatures
$y_{k}(t|n)$ (see Fig. \ref{fig:Synchronization-evolution}-a). Whilst,
in conventional synchronization nodes use as TO reference the timing
information from one not-colliding signal detected, say $\hat{\tau}_{k}\left[n\right]=\tau_{i}$,
of one neighbor node (Fig. \ref{fig:Synchronization-evolution}-b).
The TOs are periodically measured and corrected every $T_{F}$ by
comparing the local values with respect to the reference $\hat{\tau}_{k}[n]$
by the exchange of the synchronization signatures $x\left(t\right)$
among all nodes to minimize the TO mismatch one another. 

\section{Colliding vs not-Colliding Frames in Synchronization Approach\label{sec:Colliding-vs-not-Colliding}}

Synchronization at PHY-level is to align the frame structure and the
corresponding time-slots to enable MAC-level functionalities. The
timing synchronization of the network is achieved when all the $K$
frames are temporally aligned and the relative TOs are (close enough
to) zero. In uncoordinated networks, nodes randomly choose to transmit
or receive the synchronization beacon $x\left(t\right)$ to iterate
the broadcast and correction of TO several times. When the node is
transmitting, it broadcasts its synchronization state, while the receiving
nodes update their TO locally based on the information extracted from
the received signal. Thus, the $k$-th receiving node updates its
TO at $n$-th synchronization step based on the relative error $\Delta\tau_{k}\left[n\right]$
(\ref{eq:relative_sync_par}), accordingly to the updating
\begin{equation}
\begin{split}\tau_{k}\left[n+1\right] & =\tau_{k}\left[n\right]+\varepsilon\Delta\tau_{k}\left[n\right]+v_{k}\left[n\right]\\
 & =\left(1-\varepsilon\right)\tau_{k}\left[n\right]+\varepsilon\hat{\tau}_{k}\left[n\right]+v_{k}\left[n\right],
\end{split}
\label{eq:sync_upd}
\end{equation}
where $\varepsilon\in\left(0,1\right)$ is a design parameter, and
$v_{k}\left[n\right]$ is a stochastic perturbation due to oscillator's
instability. Network reaches the TO synchronization when $\tau_{1}\left[n\right]=\tau_{2}\left[n\right]=...=\tau_{K}\left[n\right]=\tau_{\infty}$
for a certain $n$, and it is achieved asymptotically ($n\rightarrow\infty$)
for a connected network. 

In dense networks (say $K\geq100$ nodes), the collision of signals
is highly likely, even more when there is no scheduler or coordinator
agent. Under this scenario, focus here is to compare two synchronization
approaches based on: collision-avoidance and collision of signals.
The former is a conventional synchronization based on master-slave
reference, where the TO reference $\hat{\tau}_{k}\left[n\right]$
is extracted from the detection of received signal free of any collision.
The latter is based on distributed synchronization where the reference
$\hat{\tau}_{k}\left[n\right]$ is estimated from the superimposition
of the received signals. Synchronization methods are thus conceptually
one the opposite of the other, and the analysis here is for collision
vs non-collision synchronization.

\subsection{Non-Collision based Synchronization \label{subsec:Not-Colliding}}

\begin{figure}[t]
\begin{centering}
\includegraphics[scale=0.5]{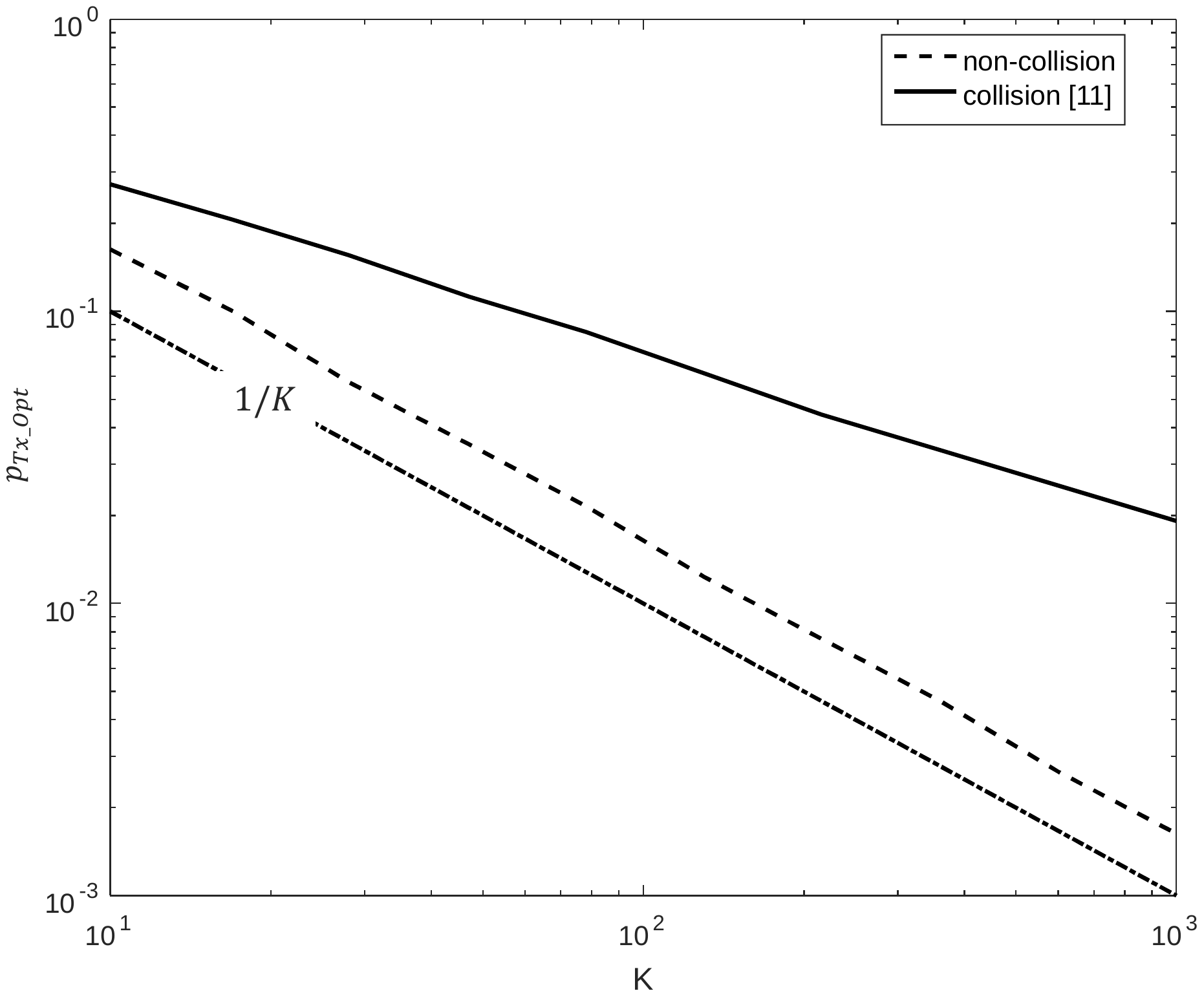}
\par\end{centering}
\caption{Optimum probability of node transmission $p_{Tx\_Opt}$ vs number
of nodes $K$ for both collision (solid) and non-collision (dash)
-based synchronization approaches. For non-collision synchronization
approach, a valid approximation of $p_{Tx\_Opt}=1/K$ is illustrated.
\label{fig:Optimum-probability}}
\end{figure}
\begin{figure*}[tp]
\begin{centering}
\includegraphics[width=18cm,height=8cm]{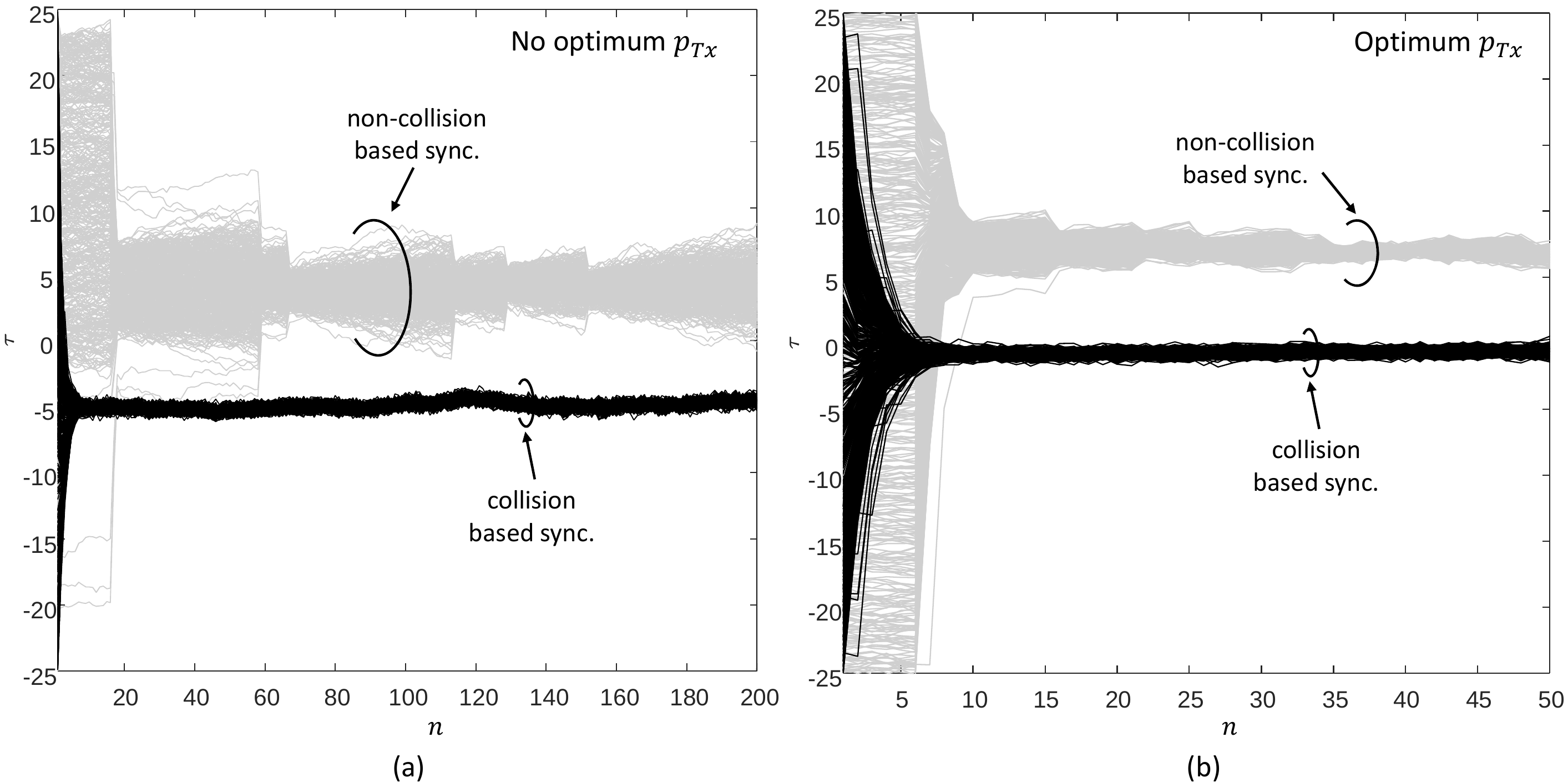}
\par\end{centering}
\caption{TO synchronization toward convergence for non-collision based synchronization
(gray lines) and collision-based distributed synchronization (black
lines), for a fully connected network of $K=500$ nodes (all-to-all
connectivity). (a) The probability of node transmission is $p_{Tx}=0.01$
for both synchronization approaches. (b) Optimum probability of transmission
for non-collision based $p_{Tx\_Opt}=0.002$ and collision-based $p_{Tx\_Opt}=0.03$.\label{fig:TOsync_vs_iter}}
\end{figure*}

The synchronization algorithm is based on master-slave reference.
Each receiver locally corrects its TO based on the relative TO error
$\Delta\tau_{k}[n]$ with respect to the TO of the signal free-of-collision
detected from its neighborhood $\mathcal{N}_{k}$, when available
(Fig. \ref{fig:Synchronization-evolution}-b). The received signal
$y_{k}\left(t|n\right)$ is a filter matched to the local copy of
the synchronization signature $x\left(t\right)$ and it is conventionally
used to search for a signal free-of-collisions. If it is detected
a signal free-of-collisions from a transmitting node, say $i$-th,
the receiver corrects its TO with respect to the reference $\hat{\tau}_{k}\left[n\right]=\tau_{i}\left[n\right]$,
otherwise the local oscillator is in free-running mode making the
TO drift-apart. 

The synchronization update (\ref{eq:sync_upd}) depends on whether
a signal free-of-collision is detected, and thus to maximize the probability
of synchronization update (or maximize the convergence rate), the
optimization problem can be addressed in term of the probability of
a device to be on transmission mode such that the probability of collision
is minimized, or equivalently the probability that only one collision-free
signal in every frame is maximized.

The probability of synchronization update (or one signal free-of-collision)
is given by
\begin{equation}
\begin{split}\text{Pr}\left\{ \text{update}|p_{Tx},K\right\} = & \text{Pr}\left\{ \text{1 collision-free signal out of}\right.\\
 & \left.\left|\mathcal{N}_{k}\right|-1\text{ colliding}\right\} \\
= & \text{Pr}\left\{ \text{1 collision-free signal}|\ell\text{ Tx nodes}\right\} \\
 & \times\text{Pr}\left\{ \ell\text{ Tx nodes out of }K\right\} ,
\end{split}
\label{eq:prob}
\end{equation}
and the optimum probability that a node transmits is given to the
optimization problem 
\begin{equation}
p_{Tx\_Opt}=\arg\max_{p_{Tx}}\text{Pr}\left\{ \text{update}|p_{Tx},K\right\} ,
\end{equation}
where the probability of synchronization update is conditioned for
a given density of nodes $K$. The optimum probability $p_{Tx}$ can
be derived analytically from model (\ref{eq:prob}) (omitted here
for sake of compactness). Figure \ref{fig:Optimum-probability} shows
this optimum probability that a node transmits $p_{Tx\_Opt}$ (dash
line) for different network connectivity or density given by varying
$K$. Increasing $K$, the synchronization algorithm can optimize
the synchronization efficiency by assigning an optimum probability
of transmission that in turn it can be approximated to $p_{Tx\_Opt}\cong1/K$.

\subsection{Collision-based Distributed Synchronization\label{subsec:Colliding}}

The distributed synchronization algorithm \cite{Alvarez2014} is based
on distributed phase locked loop (D-PLL) that is an iterative control
system for synchronization that corrects the local TO (\ref{eq:sync_upd})
on each node, say $k$-th, based on the TO error $\Delta\tau_{k}[n]$
with respect to the TO of the ensemble $\mathcal{N}_{k}$ (Fig. \ref{fig:Synchronization-evolution}-a).
The synchronization by D-PLL can be framed as consensus algorithm
where transmitting nodes broadcast their synchronization status as
TO $\tau_{i}[n]$ embedded into modulated signatures $x(t)$, and
the nodes receiving these misaligned and superimposed signatures correct
their TO accordingly. The TO reference $\hat{\tau}_{k}\left[n\right]=\sum_{i\in\mathcal{N}_{k}}a_{ki}\left[n\right]\tau_{i}\left[n\right]$
is estimated from $y_{k}\left(t|n\right)$ superimposing the ensemble
of received signatures that embed $\left\{ \tau_{i}\left[n\right]\right\} _{i\in\mathcal{N}_{i}}$. 

The estimation of the average TO $\hat{\tau}_{k}\left[n\right]$ from
the ensemble of nodes reduces to the estimation of time of delay of
superimposed multiple copies of the same signature $x(t)$ affected
by different TOs. In distributed synchronization, the problem reduces
to the estimation of TO centroids from the superposition of several
$x\left(t\right)$ in (\ref{eq:y(t)}). The estimation of the relative
TO error from collisions is based on the timing metric from the filter
matched to the synchronization signature $x\left(t\right)$
\begin{equation}
\begin{split}r\left(\tau\right)= & \int y_{k}\left(t\right)x^{*}\left(t-\tau\right)dt\\
= & \sum_{i\in\mathcal{N}_{k}}h_{ki}\left(t\right)g\left(t-\tau_{k,i}\right)\exp\left(j\Omega_{ki}\right)+\tilde{w}_{k}\left(t\right),
\end{split}
\label{eq:xcorr}
\end{equation}
where $\tilde{w}_{k}\left(t\right)=w_{k}\left(t\right)*x\left(t\right)$.
This is the superposition of $\left|\mathcal{N}_{k}\right|$ auto-correlations
$g\left(t\right)=x\left(t\right)*x^{*}\left(t\right)$ with amplitudes
$h_{ki}$ and delays
\begin{equation}
\tau_{k,i}=\tau_{i}-\tau_{k}.
\end{equation}
In distributed synchronization it is necessary to estimate the average
value of the TO of $\mathcal{N}_{k}$ (not the individual values),
and thus the barycentral delay value
\begin{equation}
\begin{split}\Delta\tau_{k} & =\frac{\int t|r(t)|^{2}dt}{\int|r(t)|^{2}dt}\simeq\frac{\sum_{i\in\mathcal{N}_{k}}|h_{ki}|^{2}\tau_{k,i}}{\sum_{i\in\mathcal{N}_{k}}|h_{ki}|^{2}}\\
 & =\left(\hat{\tau}_{k}-\tau_{k}\right)
\end{split}
\label{eq:baryc}
\end{equation}
is preferred estimator of the average delay weighted by the amplitudes
$|h_{ik}|^{2}.$ Proof of the optimality of estimator (\ref{eq:baryc})
can be shown (not here).

For an optimum duplex strategy in distributed synchronization, the
probability that a node is on transmission mode can be optimized to
maximize the convergence rate, or equivalent by minimizing the convergence
time. The decreasing rate of the TO dispersion over synchronization
iterations \cite{Alvarez2015} is
\begin{equation}
\begin{alignedat}{1} & \frac{\sigma^{2}\left[n+1\right]}{\sigma^{2}\left[n\right]}=\left(1+p_{Tx}\right)^{K}+p_{Tx}^{K}\\
 & +\sum_{\ell=1}^{K-1}\left(\frac{\ell}{K}+\frac{K-\ell}{K}\left[\left(1-\varepsilon\right)^{2}+\frac{\varepsilon^{2}}{\ell}\right]\right)\binom{K}{\ell}p^{\ell}\left(1-p\right)^{K-\ell},
\end{alignedat}
\end{equation}
where $\sigma^{2}\left[n\right]=\sum_{k,i\neq k}\left(\tau_{k}\left[n\right]-\tau_{i}\left[n\right]\right)^{2}/K\left(K-1\right)$
is the sample mean-square deviation (MSD) of the TO synchronization
from the average. Figure \ref{fig:Optimum-probability} shows also
the optimum probability of node transmission $p_{Tx\_Opt}$ (solid
line) by varying $K$ for the distributed synchronization. Since $p_{Tx\_Opt}$
for distributed synchronization is higher than the same for non-collision
method, Fig. \ref{fig:Optimum-probability} shows that distributed
method is more energy consuming.

\begin{figure}[t]
\begin{centering}
\includegraphics[scale=0.47]{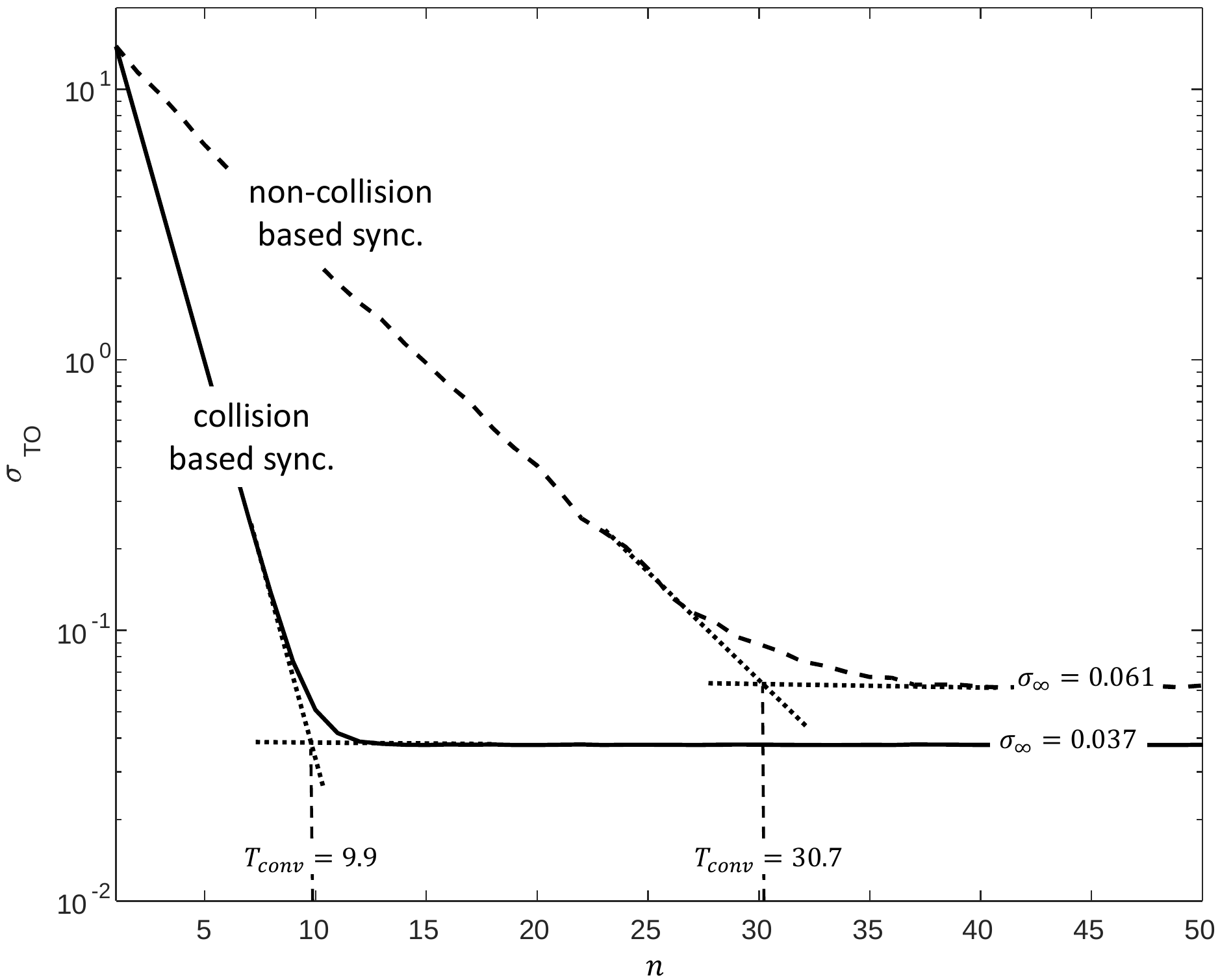}
\par\end{centering}
\caption{Root MSD $\sigma_{TO}$ vs synchronization iteration $n$ of collision
based distributed synchronization (solid line) and non-collision based
synchronization (dash line) for $K=1000$ nodes fully connected (all-to-all).
For an optimum probability of node transmission $p_{Tx\_Opt}=0.02$
(collision) and $p_{Tx\_Opt}\protect\cong1/K$ (non-collision). It
is depicted the convergence time $T_{conv}$ and root MSD at convergence
$\sigma_{\infty}$ that can be globally used to evaluate (and compare)
the performance of both synchronization approaches.\label{fig:Root-MSD_vs_iter}}
\end{figure}
\begin{figure}[t]
\begin{centering}
\includegraphics[scale=0.47]{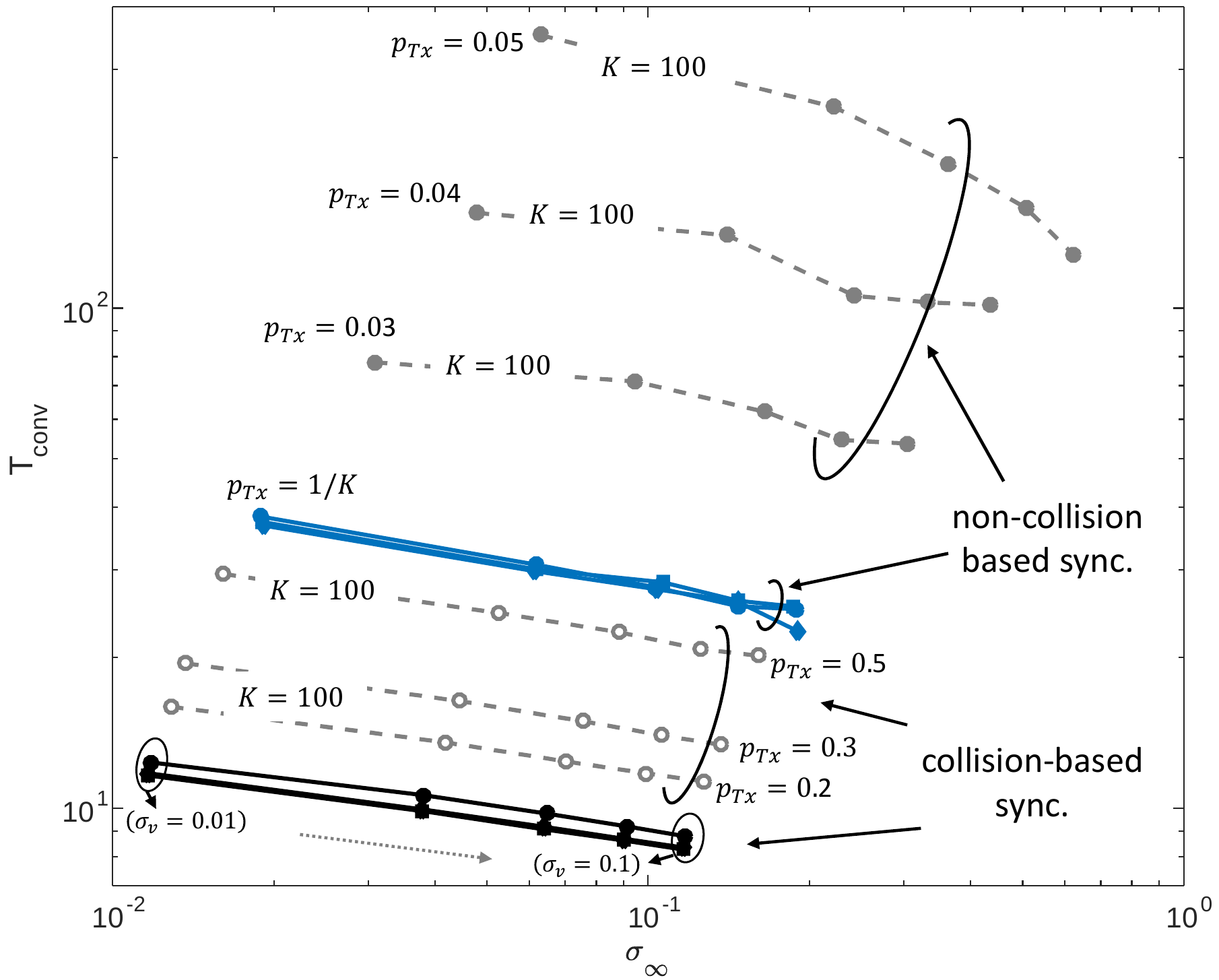}
\par\end{centering}
\caption{Convergence time $T_{conv}$ vs root MSD at convergence $\sigma_{\infty}$
(Fig. \ref{fig:Root-MSD_vs_iter} is an example) for collision-based
distributed synchronization and non-collision based synchronization
for $K=100,500,1000$ with its corresponding optimum duplexing: $p_{Tx\_Opt}=0.07,0.03,0.02$
(collision-based) and $p_{Tx\_Opt}\protect\cong1/K$ (non-collision
based). With different local-oscillator drift noise $\sigma_{v}=0.01:0.1$.
For $K=100$ is the synchronization performance for sub-optimum duplexing:
non-collision $p_{Tx}=0.03:0.05$ (filled dots), collision $p_{Tx}=0.2:0.5$
(empty dots).\label{fig:Tconv_vs_sigma_v}}
\end{figure}

\section{Numerical Results \label{sec:Numerical-Results}}

The comparative performance analysis is carried out by Monte Carlo
simulations of a dense fully-connected network that consists of $K$
nodes mutually coupled, where all nodes adopt the same signature (it
can be shown that changing signature, or implemented frequency hopping
strategies, have the effect to minimize the collision probability).
Transmission is simulated in form of frames, each frame contains the
synchronization beacon followed by payload. To set a general framework,
the discrete synchronization signature $x\left[m\right]$ can be based
on Zadoff-Chu (ZC) sequences due to their good correlation properties
\cite{Jae12}
\begin{equation}
x\left[m\right]=\exp\left(j\frac{\pi}{N}u\left(m-N_{c}\right)^{2}\right)\qquad0\leq m\leq N_{x}-1,
\end{equation}
with $N$-samples length and cyclic prefix and suffix $N_{c}$-samples
length both, for a total support $N_{x}=N+2N_{c}$, where $u$ is
the root index that is relative prime to $N$ (here $N$ is even and
$u=1$). At set-up ($n=0$) network is asynchronous: TO is normalized
by signature length $N$ and it is uniformly distributed over the
frame $\tau_{k}\left[0\right]\sim\mathcal{U}\left(-25N,25N\right)$,
and the loop gain filter is $\varepsilon=0.5$. The mean-square deviation
(MSD) $\sigma_{TO}^{2}\left[n\right]=\sum_{k,i\neq k}\left(\tau_{k}\left[n\right]-\tau_{i}\left[n\right]\right)^{2}/K\left(K-1\right)$
is used as metric for convergence dispersion of TO synchronization.

TO synchronization toward convergence is illustrated in Fig. \ref{fig:TOsync_vs_iter}
for $K=500$ nodes fully connected (all-to-all connectivity) and each
one is affected by an independent clock-drift. Fig. \ref{fig:TOsync_vs_iter}-a
compares both synchronization approaches on a not-optimum duplexing
scenario by increasing the probability of node transmission $p_{Tx}=0.01$,
for both methods. Notice that when the probability of collision increases,
the detection of a signal free-of-collision is low, and hence the
network does not reaches the minimum dispersion of TO error that allows
the proper communication. The optimum duplexing is shown in Fig. \ref{fig:TOsync_vs_iter}-b,
where $p_{Tx\_Opt}=0.002$ for no-collision and $p_{Tx\_Opt}=0.03$
collision scenario. The distributed synchronization algorithm provides
stable performance by reaching a convergence status in few iterations
and reduces the synchronization dispersion error for large $n$ on
both not-optimum and optimum duplexing. However, the fact that the
optimum $p_{Tx}$ is low in conventional synchronization, it is an
advantage on reduction of power consumption.

Figure \ref{fig:Root-MSD_vs_iter} shows the MSD $\sigma_{TO}$ vs
synchronization iteration $n$ for both collision and non-collision
-based synchronization approach for a network of $K=1000$ nodes fully
connected (all-to-all). The convergence time $T_{conv}$ and steady-state
MSD $\sigma_{\infty}$ are the metrics used to evaluate and compare
the performance of both synchronization algorithms. The optimum probability
of transmission is used for both methods to compare for the best $T_{conv}$,
that for collision-based is $p_{Tx\_Opt}=0.02$ and non-collision
is $p_{Tx\_Opt}\cong1/K$. Results in Fig. \ref{fig:Root-MSD_vs_iter}
illustrates that the distributed synchronization algorithm improves
the convergence time $T_{conv}$ by $32\%$ and the MSD $\sigma_{\infty}$
by $61\%$ in comparison to collision-avoidance scenario.

Based on the example in Fig. \ref{fig:Root-MSD_vs_iter}, the synchronization
algorithms are evaluated and compared on the basis of $T_{conv}$
and $\sigma_{\infty}$. Fig. \ref{fig:Tconv_vs_sigma_v} illustrates
the convergence time $T_{conv}$ vs MSD $\sigma_{\infty}$ for different
network density $K=100,500,1000$ with its corresponding optimum probability
of node transmission $p_{Tx\_Opt}=0.07,0.03,0.02$ (collision-based)
and $p_{Tx\_Opt}\cong1/K$ (non-collision based) by varying the perturbation
noise of the local oscillator $\sigma_{v}=0.01:0.1$. Experimental
results in Fig. \ref{fig:Tconv_vs_sigma_v} prove that the distributed
synchronization is uniformly better in terms of $T_{conv}$ and $\sigma_{\infty}$
independently of $K$ in comparison with collision-avoidance approach.
$T_{conv}$ is reduced by $30\%$, and the $\sigma_{\infty}$ is reduced
by $60\%$ for any increasing perturbation noise (due to oscillator's
instability). On the other hand, the performance of the non-collision
based synchronization is degraded significantly for any sub-optimum
probability of transmission $p_{Tx}=0.03:0.05$ thus limiting the
scalability of the network.

\section{Conclusion\label{sec:Conclusion}}

The novel distributed (collision-based) synchronization algorithm
is compared with conventional (non-collision) synchronization under
an uncoordinated dense network scenario without any reference agent
where all nodes in the network are assigned the same synchronization
signature. The conventional synchronization algorithm is based on
detection of signal free-of-collision to enable the update synchronization,
whilst distributed synchronization exploits the collision of signals
to extract the synchronization information to enable the TO update.
The impact of the network scalability is investigated for both synchronization
approaches by comparing the convergence time and synchronization dispersion
error. Numerical results prove that superposition of signals provides
remarkable improvements in synchronization compared to conventional
non-collision based synchronization, even on network settings that
degrades the performance of the synchronization. This feature makes
the proposed method more robust for synchronization in dense IoT networks.
The synchronization setting fits the requirements involved on heterogeneous
networks, allowing scalability, minimizing the signaling overhead
by using the same signature in beacon to take benefit of collision,
and considerably reducing the complexity of the synchronization algorithm
at price of a change of signal processing paradigm at PHY-layer. However,
the duplexing strategy in distributed synchronization implies that
more nodes will be on transmission mode, it translates in a drawback
of augmented power consumption.

\bibliographystyle{IEEEtran}

\end{document}